\documentclass[showpacs,amsmath,amssymb,twocolumn,pra,superscriptaddress]{revtex4-2}

\makeatletter
\newenvironment{breakablealgorithm}
{
		\begin{center}
			\refstepcounter{algorithm}
			\hrule height.8pt depth0pt \kern2pt
			\renewcommand{\caption}[2][\relax]{
				{\raggedright\textbf{\ALG@name~\thealgorithm} ##2\par}%
				\ifx\relax##1\relax 
				\addcontentsline{loa}{algorithm}{\protect\numberline{\thealgorithm}##2}%
				\else 
				\addcontentsline{loa}{algorithm}{\protect\numberline{\thealgorithm}##1}%
				\fi
				\kern2pt\hrule\kern2pt
			}
		}{
		\kern2pt\hrule\relax
	\end{center}
}
\makeatother
\allowdisplaybreaks[2]
\usepackage{algorithm}  
\usepackage{algorithmic}   
\usepackage{multirow}

\usepackage{graphicx} 
\usepackage{amsthm}
\usepackage{amsmath}
\usepackage[nopatch]{microtype}
\usepackage{amssymb}
\usepackage{braket}
\usepackage{qcircuit}
\usepackage[title]{appendix}
\usepackage{appendix}
\usepackage{titlesec}
\usepackage{titletoc}
\usepackage{float}
\usepackage{caption}
\usepackage{subcaption}
\usepackage{bm}
\usepackage{rotating}
\usepackage{algorithm}
\usepackage{algorithmic}
\usepackage{xcolor}
\usepackage{nicematrix}
\newtheorem{lemma}{Lemma}
\newtheorem{theorem}[lemma]{Theorem}
\newtheorem{definition}[lemma]{Definition} 

\newtheorem{observation}[lemma]{Observation}
\newtheorem{corollary}[lemma]{Corollary}
\newtheorem{remark}[lemma]{Remark}

\begin{document}

\title{Quantum multi-row iteration algorithm for linear systems with non-square coefficient matrices}
\author{Weitao Lin}\email{linweitao22s@ict.ac.cn}\author{Guojing Tian}\email{tianguojing@ict.ac.cn}\author{Xiaoming Sun}\email{sunxiaoming@ict.ac.cn}
\affiliation{State Key Lab of Processors,
Institute of Computing Technology, Chinese Academy of Sciences, 100190, Beijing, China}
\affiliation{University of Chinese Academy of Sciences, Beijing, 100049, China}
\date{\today}

\begin{abstract}

In the field of quantum linear system algorithms, quantum computing has realized exponential computational advantages over classical computing. However, the focus has been on square coefficient matrices, with few quantum algorithms addressing non-square matrices. Towards this kind of problems defined by \( Ax = b \) where \( A \)\( \in\mathbb{R}^{m \times n} \), we propose a quantum algorithm inspired by the classical multi-row iteration method and provide an explicit quantum circuit based on the quantum comparator and Quantum Random Access Memory (QRAM).
The time complexity of our quantum multi-row iteration algorithm is \( O(K \log m) \), with \( K \) representing the number of iteration steps, which demonstrates an exponential speedup compared to the classical version. Based on the convergence of the classical multi-row iteration algorithm, we prove that our quantum algorithm converges faster than the quantum one-row iteration algorithm presented in [Phys. Rev. A, 101, 022322 (2020)]. Moreover, our algorithm places less demand on the coefficient matrix, making it suitable for solving inconsistent systems and quadratic optimization problems.
\end{abstract}

\maketitle

\section{Introduction}

Quantum Linear System Algorithms (QLSAs) are widely used solvers in quantum computation, 
which play essential roles in solving problems in finance~\cite{kubo2021variational,an2021quantum}, bio-computing~\cite{lambert2013quantum}, fluid dynamics~\cite{chen2021quantum}, machine learning~\cite{biamonte2017quantum}, etc. 
Since Harrow, Hassidim and Lloyd's pioneering presentation of the first quantum linear system solver~\cite{harrow2009quantum}, QLSA has been proved to have an exponential acceleration over the classical one on the dependence of the problem size. 
Besides the exponential acceleration achieved from the presentation of quantum states, much effort has been devoted to improving the dependence on the condition number, sparsity and accuracy~\cite{childs2017quantum,wossnig2018quantum,subacsi2019quantum,lin2020optimal,an2022quantum,costa2022optimal}. The quantum algorithm with $O(\kappa\log(1/\epsilon))$ complexity for solving linear systems in~\cite{costa2022optimal} is asymptotically optimal, where $\kappa$ is the condition number and $\epsilon$ is the error tolerance.

The above results mainly focus on the linear equations with square coefficient matrix, i.e., the number of constraints (rows) and variables (columns) are equal,
thus there exists one unique solution.  
However, when the coefficient matrices are not square, the unique solutions may not exist. We refer to this kind of linear system as ELS (Extended Linear System) in the subsequent of this article. ELS problems are commonly seen in areas such as image processing~\cite{andersen1989algebraic}, machine learning~\cite{suykens1999least}, and computed tomography~\cite{gordon1970algebraic}. 
Under these circumstances, almost all existed methods could not work.
Firstly, the QLSAs mentioned above, e.g. solving the inverse matrices, are unsuitable because solving the Moore-Pseudo inverse differs from solving an inverse matrix. 
Secondly, in fact, in classical computation theory, there are some standard methods for solving ELS problems, such as the QR decomposition (Orthogonal matrix and Upper triangular matrix decomposition)~\cite{gander1980algorithms}, the SVD (Single Value Decomposition)~\cite{wall2003singular}, the iteration methods~\cite{strohmer2009randomized} and so on. The time complexity of the QR decomposition and the SVD is $O(mn^2)$, where $m$ represents the number of rows of the coefficient matrix and $n$ means the number of columns. The time complexity for the iteration methods is $O(Kn)$, where $K$ is the number of iteration steps that rely on the specific iteration methods, which explicitly or implicitly depend on $m$. As the dimension of the coefficient matrix increases, it becomes more complex and more costly to solve ELS problems.

So the natural question is: can quantum computation accelerate the process of solving these ELS problems? Two avenues may be feasible. Firstly, we may consider transforming the non-square matrix into a square one and applying the QLSA to solve it. But we need to add more restrictions to the linear systems to achieve this. For example, we can introduce a square symmetric matrix $A'=\left(\begin{array}{cc}
0&A  \\A^T&0 \end{array}\right)$ when solving equations $A \bm x= \bm b$ with $A\in\mathbb{R}^{m\times n}$. However, $A'$ is not full rank. Therefore, it is not invertible, which leads to a failure to solve it through solving the inverse of $A'$.
Then, we consider constructing a square and invertible matrix from $A$. One choice is $(A^TA)^{-1}$ and the solution will be $x=(A^TA)^{-1}A^Tb$, but it needs $A$ to be column-full-rank.
While this idea has been employed in~\cite{wu2021quantum}, they focus on designing a {quantum-classical hybrid} algorithm instead of a pure quantum one. 
Therefore, transforming the non-square matrix to a square one may not be an effective way to solve ELS problems.

The second avenue is to consider the quantum iteration method.
Some QLSAs have the potential to solve the ELS problems through this idea~\cite{kerenidis2020quantum,shao2020row}. 
The quantum one-row iteration method~\cite{shao2020row}, the quantum version of the Kaczmarz (one-row) iteration method~\cite{Kaczmarz1937}, is a viable technique to directly solve ELS, though they focus on the problem with the coefficient matrix being square. The running time of their algorithm is $O(\kappa_s^2(A)log\frac{n}{\epsilon})$, where $\kappa_s(A)=\frac{\|A\|_F}{\|A^{-1}\|}$, $\|A\|_F$ is the frobenius norm of the matrix A, $\|A^{-1}\|$ is the norm of $A^{-1}$, $n$ is the size of the problem and $\epsilon$ is the error tolerance. The algorithm exhibits an exponential speedup over the Randomized Kaczmarz algorithm (classical one-row iteration algorithm). Yet, the convergence rate remains similar to the classical one-row Randomized Kaczmarz algorithm. 
Fortunately, we found the classical multi-row iterations in~\cite{lian2015asynchronous,moorman2021randomized} have demonstrated quicker convergence rates for solving linear systems with both square and non-square coefficient matrices. Naturally, we consider whether the quantum version of the multi-row methods leads to a higher convergence rate while keeping the exponential speedup.

In this paper, we propose an approach that solves linear equations with a coefficient matrix of size $m\times n,m\ge n$, using the idea of LCU (linear combinations of unitaries~\cite{childs2017quantum}). We design a quantum multi-row iteration algorithm with an explicit quantum circuit. The quantum circuit builds on the quantum comparator and QRAM (Quantum Random Access Memory), and can complete multi-row iteration with different iteration weights. The gates used for one iteration step scale logarithmic on the problem size. Our quantum multi-row iteration algorithm converges faster than the quantum one-row iteration algorithm~\cite{shao2020row} and keeps the same exponential speedup. This acceleration will yield considerable advantages in large-scale problems. Besides, since the convergence rate improvement is controlled by $\frac{\alpha_A^2}{q}$, where $\alpha_A$ is the relaxation parameter and $q$ is the number of rows chosen, increasing $\alpha_A$ to a critical point and increasing $q$ will not only significantly improve the convergence rate but also achieve a better convergence horizon. Moreover, regardless of the existence of the solution of the system, the solver will return an exact solution in the consistent systems (systems with a unique solution) or the least square solution in the inconsistent systems (systems without a unique solution). The solver can also be used as a stochastic gradient descent iterator for some specific loss functions.

The outline of the paper is as follows. Sec.\ref{Preliminari} gives some preliminaries including the classical multi-row iteration method, the quantum one-row iteration, and block-encoding. In Sec.\ref{Algorithm}, we present our quantum multi-row algorithm. We first show the key points of the algorithm. Then, we define the date structure for efficient state preparation and finally the sketch of the algorithm. Analysis for the resource is provided in the Sec.\ref{anan}. Numerical experiment is shown in Sec.\ref{Numerical experiment}. A brief illustration of the application is presented in Sec.\ref{Application}. 

\section{Preliminaries}\label{Preliminari}
In this section, we will present some existing significant results including the classical multi-row iteration method~\cite{moorman2021randomized} and the quantum one-row iteration approach~\cite{shao2020row}. Next we will briefly introduce block-encoding, which is a technique of embedding a non-unitary matrix into a large unitary one. The explicit quantum circuit for block-encoding will be postponed to
the next section. 

\subsection{Classical iteration method}\label{class_one}
In classical computation theory, the randomized one-row iteration method is usually used to solve linear systems, especially when the coefficient matrix is non-square~\cite{Kaczmarz1937}.
Given $A\in\mathbb{R}^{m\times n}$ and $\bm b\in\mathbb{R}^m$, the purpose of iteration methods is to find $\bm x\in\mathbb{R}^n$ which satisfies the linear system of equations
\begin{equation}
    A \bm x=\bm b.
\end{equation}
Otherwise, we define the least-square solution  
\begin{equation}
    \bm x^*:=\mathop{\arg\min}\limits_{x\in\mathbb{R}^n}\frac{1}{2}\|\bm b-A \bm x\|^2,
\end{equation}
where $\|\cdot\|$ denotes the 2-norm.

The randomized Kaczmarz iteration method~\cite{gordon1970algebraic,Kaczmarz1937} gives the following iteration scheme
\begin{equation}\label{one_row}
    \bm x^{k+1}=\bm x^k+\frac{ b_{i_k}-A_{i_k} \bm x^k}{\|A_{i_k}\|^2} A_{i_k}^T
\end{equation}
where $b_{i_k}$ is the $i_k$-th element of the vector $\bm b$ and $A_{i_k}$ is the $i_k$-th row of the matrix $A$. This method is an alternating projection method. That is, $\bm x^{k+1}$ is the orthogonal projection of $\bm x^k$ onto the hyperplane $A_{i_k}\bm x=b_{i_k}$, {via which it continuously approximates the exact solution by alternating projections}.  
The randomized one-row iteration has been proven to have exponential convergence rates~\cite{strohmer2009randomized} and is generalized to solve inconsistent systems~\cite{needell2010randomized}. Many studies have attempted to enhance the rate of convergence for this approach~\cite{lian2015asynchronous,liu2016accelerated,necoara2019faster,xiao2023fast}, including the randomized multi-row iteration method~\cite{moorman2021randomized}.

The classical multi-row iteration method~\cite{moorman2021randomized} gives the following iteration strategy.
\begin{lemma}[\cite{moorman2021randomized}]\label{multistra} 
Given a linear system of equations $A\bm x=\bm b$, where $A\in\mathbb{R}^{m\times n},m\ge n$ and $\bm b\in\mathbb{R}^m$. Then, there exists a multi-row iteration protocol
\begin{equation}\label{multi}
    \bm x^{k+1}=\bm x^k+\frac{1}{q}\sum\limits_{i\in\tau_k}\omega_i\frac{b_i-A_i \bm x^k}{\|A_i\|^2}A_i^T
\end{equation}
where $\tau_k$ is a random set of $q$ row indices sampled with replacement and $\omega_i$ represents the weight corresponding to the $i$th row.
\end{lemma}

Let $\bm x^*$ be the exact solution or least square solution of $A\bm x=\bm b$ and $\bm e^k=\bm x^k-\bm x^*$ be the iteration error of step $k$. We introduce the following definition to help show the convergence rate.
\begin{definition}\label{deff}
    Let $Diag(d_1,d_2,\cdots,d_m)$ denote the diagonal matrix with $d_1,d_2,\cdots,d_m$ on the diagonal. Define the normalization matrix
    \begin{equation}
        D:=Diag(\|A_1\|,\|A_2\|,\cdots,\|A_m\|)
    \end{equation}
such that the matrix $D^{-1}A$ has rows with unit norm, the probability matrix
    \begin{equation}
        P:=Diag(p_1,p_2,\cdots,p_m)
    \end{equation}
where $p_i$ denotes the probability of choosing the i-th row, and the weight matrix
\begin{equation}
    W:=Diag(\omega_1,\omega_2,\cdots,\omega_m)
\end{equation}
\end{definition}

Thus, the convergence rate of Eq.(\ref{multi}) can be given as follows.
\begin{lemma}[\cite{moorman2021randomized} Theorem 1]
\label{lemma2}
 Given the iteration strategy defined in Lemma.\ref{multistra} and suppose the matrix $P$ and $W$ in Definition.\ref{deff} are chosen such that $PWD^{-2}=\frac{\alpha_A}{\|A\|_F^2}I$, the convergence rate satisfies
    \begin{equation}
    \begin{aligned}
        \mathbb{E}&[\|\bm e^{k+1}\|^2]\\&\le\sigma_{max}\left((I-\alpha_A\frac{A^TA}{\|A\|_F^2})^2-\frac{\alpha_A^2}{q}(\frac{A^TA}{\|A\|_F^2})^2\right)\|\bm e^k\|^2\\&+\frac{\alpha_A}{q}\frac{\|\bm r^k\|_W^2}{\|A\|_F^2} 
    \end{aligned}
\end{equation}
where $\sigma_{max}$ is the maximum singular value, $\|A\|_F^2=\sum_{i,j}A_{ij}^2$, $\alpha_A>0$ is the relaxation parameter, $\|\cdot\|_W^2=\braket{\cdot,W}$, and $\bm r^k:=\bm b-A\bm x^k$ is the residual of the $k$th iteration. 
\end{lemma}

If the condition $PWD^{-2}=\frac{\alpha_A}{\|A\|_F^2}I$ is satisfied and $q$ goes to infinity, the error will converge to zero. 
If the above condition is not satisfied, the iteration result will approach a weighted least square solution instead of the least square solution itself~\cite{moorman2021randomized}. 


\subsection{Quantum one-row iteration method}\label{quantum_one_row}
The quantum version of the one-row iteration method is given in~\cite{shao2020row}. Set $\bm x^k=\|\bm x^k\|\ket{x^k}$ and $A_{i_k}=\|A_{i_k}\|\ket{A_{i_k}}$, then, the Eq.(\ref{one_row}) can be rewritten as
\begin{equation}
    \ket{x^{k+1}} = \|\bm x^k\|(I-\ket{A_{i_k}}\bra{A_{i_k}})\ket{x^k}+\frac{b_{i_k}}{\|A_{i_k}\|}\ket{A_{i_k}},
\end{equation}
up to a global phase.

Intuitively, one can define the following unitary as the iteration operator (assuming $\|A_{i_k}\|=1$), which is the basic idea of~\cite{shao2020row},
\begin{equation}\label{one_matrix}
\begin{aligned}
    U&_k=\left[
     \begin{array}{cc}
         I-\ket{A_{i_k}}\bra{A_{i_k}} & \ket{A_{i_k}}\bra{A_{i_k}} \\
         \ket{A_{i_k}}\bra{A_{i_k}} & I-\ket{A_{i_k}}\bra{A_{i_k}}
     \end{array}
    \right]\\
    &= \big(I_2\otimes V_{i_k}\big)\big(I_2\otimes(I-\ket{0}\bra{0})+X\otimes\ket{0}\bra{0}\big)\big(I_2\otimes V_{i_k}^\dagger\big)
\end{aligned}
\end{equation}
where $V_{i_k}$ represents the state preparation process $V_{i_k}\ket{0}=\ket{A_{i_k}}$, which can be achieved by the access to QRAM or by a state preparation operator. Applying the operator $U_k$ on the state $\frac{\|x^k\|}{c}\ket{0}\ket{\bm x^k}+\frac{b_{i_k}}{c}\ket{1}\ket{A_{i_k}}$, where $c$ is a normalization factor, can complete an iteration step. Similarly,  one may employ a comparable approach in formulating the iteration operators for the multi-row iteration method. However, the multi-row iteration operator constructed this way is not unitary. The tricky part lies in how to tackle this problem. In the forthcoming sections, we will demonstrate how to realize the non-unitary operator in the multi-row case. 

\subsection{Block-encoding}
To perform the non-unitary operator, we will use the technique of block-encoding, which embeds the non-unitary operator into a large unitary one. The idea of block-encoding~\cite{low2017optimal,chakraborty2018power} is widely used in the quantum algorithms associated with matrix multiplication~\cite{gilyen2019quantum} and Hamiltonian simulation~\cite{low2017optimal}. 
\begin{definition}[\cite{gilyen2019quantum} Block-encoding] 
Assume $A$ is an $s-qubit$ operator, $\alpha,\epsilon\in\mathbb{R}^+$ and $a\in\mathbb{N}$. Then the $(s+a)-qubit$ unitary operator $U$ is an $(\alpha;a;\epsilon)$-block-encoding of $A$, if
    \begin{equation}
        \left\|A-\alpha(\bra{0}^{\otimes a}\otimes I)U(\ket{0}^{\otimes a}\otimes I)\right\|\le\epsilon,
    \end{equation}
\noindent where the norm is the 2-norm. The parameters $\alpha$ and $a$ are, respectively, the subnormalization factor of the matrix and the number of ancilla qubits used. Since $\|U\|=1$, therefore $\|A\|\le \alpha$.
\end{definition}

Every unitary operator is already a (1;0;$\epsilon$)-block-encoding, and a non-unitary operator $A$ can be embedded in a ($\|A\|$;a;$\epsilon$)-block-encoding. Given a block-encoding $U$, one can prepare the state $\frac{A\ket{\psi}}{\|A\ket{\psi}\|}$ from an initial state $\ket{0}\ket{\psi}$, that is, $U \ket0 \ket \psi = \ket{0} \frac{A\ket{\psi}}{\|A\ket{\psi}\|}$. This concept enhances the applicability of quantum linear algebra, thus making it more widely applicable.
In this paper, based on the existence of the unitary operator $U$, we further present a specific construction of $U$ by decomposing it into the quantum elementary gates.

\section{Quantum multi-row iteration algorithm}\label{Algorithm}

In this section, we show several problems needed to be conquered before designing the explicit quantum circuit for the quantum multi-row iteration algorithm and provide a sketch of our algorithm. This section is organized as follows. In Sec.\ref{Key points}, we show the main difficulties and the methods to solve them. In Sec.\ref{data}, we define the data structure which is used as an efficient state preparation process in the algorithm. The sketch of the algorithm is given in Sec.\ref{The sketch of the algorithm}.

\subsection{Key points of the algorithm implementation}\label{Key points}

In this part, we will clarify the key points of the algorithm implementation, which conquer the main barrier of designing the algorithm. The barrier is the non-unitarity of the matrix corresponding to the idea of the quantum one-row iteration and we should design a proper method to implement the non-unitary operator.

\subsubsection{Non-unitarity of the matrix}

Suppose the coefficient matrix is $A\in\mathbb{R}^{m\times n}$. Based on Eq.(\ref{multi}), set $\bm x^k=\|\bm x^k\|\ket{x^k}$ and $A_i=\|A_i\|\ket{A_i}$, we can derive the following formula
\begin{equation}\label{OOOOSOSO}
\begin{aligned}
    \ket{x^{k+1}}=&\|\bm x^k\|\left(I-\frac{1}{q}\sum_{i\in\tau_k}\omega_i\ket{A_i}\bra{A_i}\right)\ket{x^k}\\&+\frac{1}{q}\sum_{i\in\tau_k}\frac{\omega_ib_i}{\|A_i\|}\ket{A_i},
\end{aligned}
\end{equation}
up to a global phase, where $A_i$ is the vector of the $ith$ row.
This is one iteration step from $k$ to $k+1$. The analog of the iteration operator shown in Eq.(\ref{one_matrix}) in the quantum multi-row iterations is given as follows 
\begin{equation}
    T_k=\left[
    \begin{array}{cc}
       I-\frac{1}{q}\sum_{i\in\tau_k}\omega_i\ket{A_i}\bra{A_i}  & \frac{1}{q}\sum_{i\in\tau_k}\omega_i\ket{A_i}\bra{A_i} \\
       \frac{1}{q}\sum_{i\in\tau_k}\omega_i\ket{A_i}\bra{A_i}  & I-\frac{1}{q}\sum_{i\in\tau_k}\omega_i\ket{A_i}\bra{A_i}
    \end{array}
    \right]
\end{equation}
Unfortunately, unlike the matrix given in Eq.(\ref{one_matrix}),  $T_k$ is not unitary. We show this non-unitarity through calculating the upper left corner of $T_kT_k^\dagger$, in which we assume that all the elements are real numbers for simplicity.
\begin{equation}
\begin{aligned}
    &\left(I-\frac{1}{q}\sum_{i\in\tau_k}\omega_i\ket{A_i}\bra{A_i}\right)^2+\left(\frac{1}{q}\sum_{i\in\tau_k}\omega_i\ket{A_i}\bra{A_i}\right)^2\\
    =&
    I-\frac{2}{q}\sum_{i\in\tau_k}\omega_i\ket{A_i}\bra{A_i}+2\left(\frac{1}{q}\sum_{i\in\tau_k}\omega_i\ket{A_i}\bra{A_i}\right)^2\\
    \neq & I
\end{aligned}
\end{equation}
The rest part of $T_kT_k^\dagger$ can be similarly computed. They all indicate that $T_k$ is not a unitary operator.
There are two reasons for this non-unitarity. First, the row vectors of matrix $A$ are not orthogonal to each other; second, the iteration weight of each iteration row is not necessarily one.

\subsubsection{Overcome the non-unitarity}\label{Overcome the non-unitarity}

Causes for the non-unitarity are already given, i.e., non-orthogonality of rows and non-unity of weight, next we will overcome them.

{\textit{1. Orthogonality.}-}
We consider to introduce ancillary qubits to generate orthogonality. The state $\ket{x^k}$ is replaced by $\sum_{i\in\tau_k}s_{k,i}\ket{i}\ket{x^k}$, where $s_{k,i}$ represents the amplitude labelled by $i$. This term is the result of the previous $(k-1)th$ iteration step after exchanging the index set from $\tau_{k-1}$ to $\tau_k$. To be consistent with Eq.(\ref{OOOOSOSO}), we use $\omega_{k,i}$ as a modified weight term, which represents $\frac{\omega_{i}}{q}$. We use $s_{k+1,i}'$ to represent the weight of the iteration result before changing the index set from $\tau_{k}$ to $\tau_{k+1}$. Then we derive the iteration step with ancillary qubits below.
\begin{equation}\label{rerererer}
\begin{aligned}
\sum_{i\in\tau_k}s_{k+1,i}'\ket{i}\ket{x^{k+1}} & =  \|\bm x^k\| \left(I-\Sigma_{\tau_k}\right)\sum_{i\in\tau_k}s_{k,i}\ket{i}\ket{x^k}\\
 & +  \sum_{i\in\tau_k}\frac{\omega_{k,i}b_i}{\|A_{i}\|}s_{k,i}\ket{i}\ket{A_{i}},
\end{aligned}
\end{equation}
where $\Sigma_{\tau_k}=\sum_{i\in\tau_k}\omega_{k,i}(\ket{i}\bra{i}\otimes \ket{A_i}\bra{A_i})$.
A simple pre-processing procedure can obtain this rescaled weight $\omega_{k,i}$, as all the weights are artificially chosen. 
The orthogonality of $\ket{i}$ makes the non-orthogonality of rows of $A$ to be orthogonal.
To cancel the effect of the norm $\|A_{i}\|$, we need to perform a pre-processing procedure to update $b_i$ as $\frac{b_i}{\|A_i\|}$ to achieve the same effect.

{\textit{2. Weight.}-}
The usage of the index register $\ket{i}$ reduces the difficulty of directly attaching the iteration weight to $\ket{A_i}\bra{A_i}$. We may consider to achieve $\sum_{i\in\tau_k}\omega_{k,i}\ket{i}\bra{i}$ instead. We intend to apply the following operator,
\begin{equation}\label{intend1}
\begin{aligned}
    \Big(\frac{\omega_{k,i}}{2}\ket{j}\bra{j}+&\frac{\omega_{k,i}}{2}\ket{j+m}\bra{j+m}\Big)\otimes(\ket{i}\bra{i}),\\ &j=i,i\in\tau_k
\end{aligned}
\end{equation}
If $\sum_{i\in\tau_k}\omega_{k,i}=1$, a state preparation process and its inverse are enough. However, in the general case  $\sum_{i\in\tau_k}\omega_{k,i} \neq 1$. 
Applying a simple state preparation process will result in some redundant parts.
To eliminate the influence of the redundant parts, we apply the following operator,
\begin{equation}\label{intend2}
\begin{aligned}
    \Big(r_{k,i}\ket{j}\bra{j}-&r_{k,i}\ket{j+m}\bra{j+m}\Big)\otimes(\ket{i}\bra{i}),\\&j=i,i\notin\tau_k
\end{aligned}
\end{equation}
where $m$ is the number of rows of the matrix $A$ and $r_{k,i}$ represents the amplitude of the redundant states. Applying the above operators on the state $\sum_i\sum_i\ket{j}\ket{i}$ has the same effect as applying $\sum_{i\in\tau_k}\omega_{k,i}\ket{i}\bra{i}$ on the state $\sum_i\ket{i}$. 

Introducing orthogonality and using the block-encoding to apply the weight, we can therefore apply the following iteration matrix $U_k$,
\begin{equation}\label{Uk}
    U_k=\left[
    \begin{array}{cc}
        I-\Sigma_{\tau_k} & \Sigma_{\tau_k} \\
         \Sigma_{\tau_k} & I-\Sigma_{\tau_k} 
    \end{array}
    \right]\\
\end{equation}
where $\Sigma_{\tau_k}=\sum_{i\in\tau_k}\omega_{k,i}(\ket{i}\bra{i}\otimes \ket{A_i}\bra{A_i})$.
It should be noted that we also use ``$U_k$'' denoting the multi-row iteration operator.

\subsubsection{Equivalent implementation}\label{Equivalent implementation}

The problem remained is how to efficiently implement Eq.(\ref{intend1}) and (\ref{intend2}) by quantum circuits. Our idea is to consider the application of the weight factor and the process of selecting the desired index $\ket{i}$ separately. Specifically, we rewrite the procedure into the block-encoding form,
\begin{equation}\label{ttototo}
 \braket{0|(G\otimes I)\tilde{U}_k(G^\dagger\otimes I)|0}
\end{equation}
where $G$ is a state prepare operator,

\begin{equation}\label{mu}
\begin{aligned}
G\ket{0}&=\sum_{\substack{j\in\tau_k,\\j\in [m]}}\sqrt{\frac{\omega_{k,j}}{2}}\ket{j}+\sum_{\substack{(j-m)\in\tau_k,\\j\in[2m]/[m]}}\sqrt{\frac{-\omega_{k,j-m}}{2}}\ket{j}\\&+\sum_{\substack{j\notin\tau_k,\\j\in [m]}}\sqrt{r_{k,j}}\ket{j}+\sum_{\substack{(j-m)\notin\tau_k,\\j\in[2m]/[m]}}\sqrt{-r_{k,j-m}}\ket{j},
\end{aligned}
\end{equation}
with $[m]=\{0,1,2,\cdots,m-1\}$. And $\tilde{U}_k$ is a linear combination of unitary
\begin{equation}
\begin{aligned}
    \tilde{U}_k&=\sum^{m-1}_{j=0}\ket{j}\bra{j}\otimes C^{(j,j)}_1+\sum^{2m-1}_{\substack{j=m\\j-m\in\tau_k}}\ket{j}\bra{j}\otimes C^{(j-m,j-m)}_{-1}\\
    &+\sum^{2m-1}_{\substack{j=m\\j-m\notin\tau_k}}\ket{j}\bra{j}\otimes C^{(j-m,j-m)}_{1}
\end{aligned}
\end{equation}
where $C_h^{(j,j)}, h\in \{1,-1\}$ is defined as 
    \begin{equation}
    \begin{aligned}
        \left(\sum^{m-1}_{l=0}\ket{(2j-l)\mod m}\bra{l}\right)\left(\sum_{l\neq j}\ket{l}\bra{l}+h\ket{j}\bra{j}\right)
    \end{aligned}
    \end{equation}
It is obvious that $C^{(j,j)}_h$ is unitary when $h=1$ or $h=-1$. $\frac{1}{2}(C_{1}^{(i,i)}-C_{-1}^{(i,i)})$ represents a matrix with the $i$th diagonal element being 1, which is $\ket{i}\bra{i}$, and $\frac{1}{2}(C_{1}^{(i,i)}-C_{1}^{(i,i)})$ is a matrix with all elements being 0. The main point of our idea is to apply an operator equivalent to the following linear combination,
\begin{equation}
    \begin{aligned}
    \sum_{i\in\tau_k}\omega_{k,i}\ket{i}\bra{i}&=\sum_{i\in\tau_k}(\frac{\omega_{k,i}}{2}C_1^{(i,i)}-\frac{\omega_{k,i}}{2}C_{-1}^{(i,i)})\\&+\sum_{i\notin\tau_k}(\frac{r_{k,i}}{2}C_1^{(i,i)}-\frac{r_{k,i}}{2}C_{1}^{(i,i)}),
    \end{aligned}
\end{equation}

To give an explicit circuit for this block-encoding, we should consider how to implement the linear combination of unitary $\tilde{U}_k$, since $G$ is a state preparation operator. We can use the idea of equivalent implementation to achieve this, which is inspired by~\cite{wan2021block}.

The application of $\tilde{U}_k$ on the basis state is given as 
\begin{widetext}  
        \begin{equation}\label{uuuuuk}
        \tilde{U}_k\ket{j}\ket{l}=\left\{
        \begin{array}{ll}
        \ket{j}\ket{(2j-l)\mod m},&0\le j\le m-1\\
        \ket{j}\ket{(2j-l)\mod m},&m\le j\le 2m-1\\
        -\ket{j}\ket{(2j-l)\mod m},&m\le j\le 2m-1,l=(j\mod m),j-m\in\tau_k
        \end{array}
        \right.
    \end{equation}
\end{widetext}
This can be equivalently implemented by a process $U_{eq}$, which contains a quantum comparator~\cite{cuccaro2004new,li2020efficient}, a quantum modular adder~\cite{cuccaro2004new,li2020efficient} and a NOT gate controlled by the comparison results. The comparator selects the states which meet the condition and the modular adder is used to prepare $\ket{(2j-l)\mod m}$. Introducing an ancilla state, which is initialized in $\ket{-}=\frac{\ket{0}-\ket{1}}{\sqrt{2}}$, the NOT gate will be applied to this state when the condition $m\le j\le 2m-1,l=(j\mod m),j-m\in\tau_k$ is satisfied. The action of the NOT gate is equivalent to multiplying $-1$. More details are given in Appendix.\ref{UEQ}.

\subsection{Data structure}\label{data}

We can correspondingly define an iteration matrix $U_k$ from the iteration process is given in Eq.(\ref{rerererer}).
This operator can complete 
the $k$th iteration through operating on $\beta_k\ket 0 \sum_{i\in\tau_k}s_{k,i}\ket{i} \ket{x^{k}}+\ket 1 \sum_{i\in\tau_k}\gamma_{k,i}s_{k,i}\ket{i}\ket{A_i} $, where $\beta_k$ and $\gamma_{k,i}$ are factors associated with $\|\bm x^k\|$ and $b_i$. Similar to Eq.(\ref{one_matrix}), we rewrite the matrix as
\begin{equation}\label{UUk}
\begin{aligned}
    U_k=&I_2\otimes\big(I-\tilde{V}\sum_{i\in\tau_k}\omega_{k,i}\ket{i}\ket{0}\bra{0}\bra{i}\tilde{V}^\dagger\big)\\+&X\otimes\big(\tilde{V}\sum_{i\in\tau_k}\omega_{k,i}\ket{i}\ket{0}\bra{0}\bra{i}\tilde{V}^\dagger\big)
\end{aligned}
\end{equation}
where the implementation of $\sum_{i\in\tau_k}\omega_{k,i}\ket{i}\bra{i}$ is achieved by $\braket{0|(G\otimes I)\tilde{U}_k(G^\dagger\otimes I)|0}$ and $\tilde{V}\sum_i\ket{i}\ket{0}=\sum_i\ket{i}\ket{A_i}$.

Besides the implementation of $\tilde{U}_k$, the efficiency of the algorithm relies on the state preparation processes $G$ and $\tilde{V}$. To efficiently prepare the row vectors and the weight vectors, we introduce a data structure as QRAM (Quantum Random Access Memory), which is different from the one in~\cite{giovannetti2008quantum}.

This data structure can be visited by an algorithm that outputs the quantum state $\ket{A_i}, A_i\in\mathbb{R}^{n}$ corresponding to the $i$th row $A_i$ of the matrix $A$, and the quantum state $\ket{\bm \omega}, \bm \omega\in\mathbb{R}^{2m}$ corresponding to the weight vector with the redundant part.
Specifically, we denote the procedures as unitary operators $V$ and $G$, where $V\ket{i}\ket{0}=\ket{i}\ket{A_i}$ and $G\ket{0}=\ket{\bm \omega}$. And we denote the multi-row readout process as $\tilde{V}\sum_i\ket{i}\ket{0}=\sum_i\ket{i}\ket{A_i}$.

\begin{figure}[H]
    \centering
    \includegraphics[width=\linewidth]{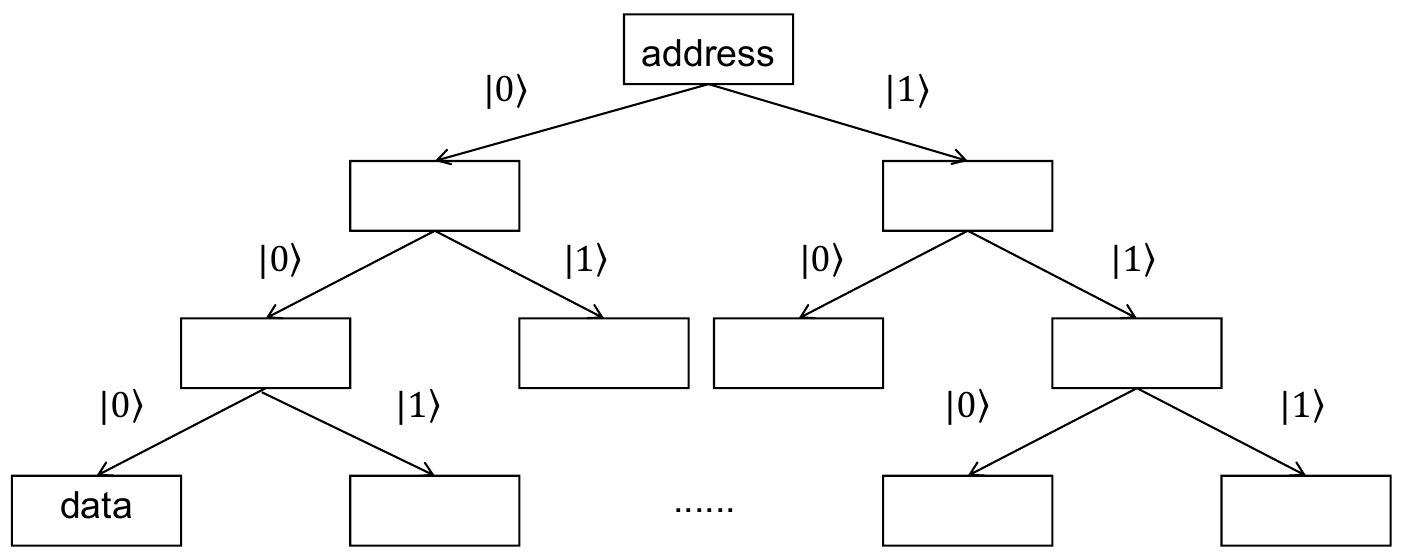}
    \caption{Schematic diagram of data structure. The root gets an address as an input and finds the routes to the corresponding leaves based on each qubit of the address. Each leaf points to a data tree, which stores the row vector of the matrix $A$.}
    \label{memory}
\end{figure}

\begin{figure}[H]
    \centering
    \includegraphics[width=\linewidth]{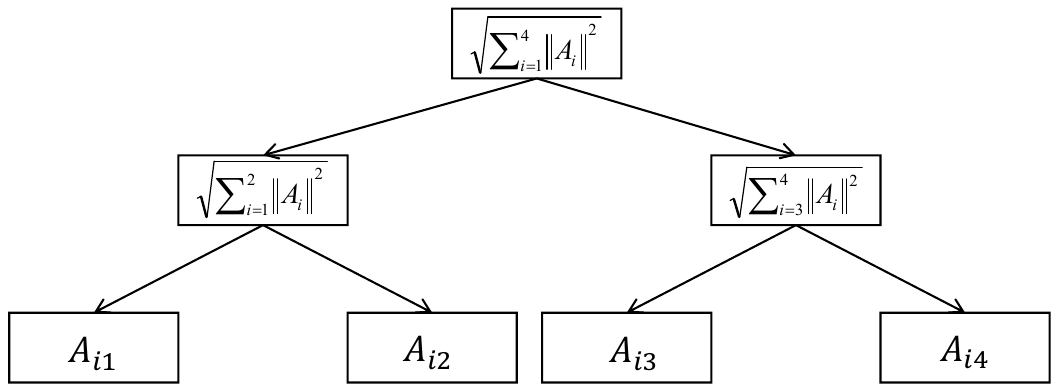}
    \caption{An example of the data tree with $n=4$. The leaves hold the individual amplitudes of the elements of the vector, and each internal node holds the square root of the sum of the squares of the norm for the value in children nodes.}
    \label{datatree}
\end{figure}

We introduce two kinds of binary trees for the data structure: one is the address tree, and the other is the memory tree.
The address tree possesses $m$ leaves, and each leaf stores the address (see Fig.\ref{memory}). 
Accessing the address tree can be analogized to a routing process. The corresponding memory tree will be activated once the addressing qubits have been set.
There are two types of memory trees: one with $n$ leaves called data tree, which stores the row vector, and the other with $2m$ leaves called weight tree, which stores the weight vector. $m$ data trees accompanied with one address tree accomplish the process $\tilde{V}\sum_i\ket{i}\ket{0}=\sum_i\ket{i}\ket{A_i}$ and the weight tree achieves $G\ket{0}=\ket{\bm \omega}$.
The leaves of the data trees and weight tree hold the individual amplitudes of the vector, and each internal node holds the square root of the sum of the squares of the norm for the value in children nodes (see Fig.\ref{datatree} as an example).
For a single reading procedure, when the binary tree gets an address as input, which can be a superposition state, the data structure finds the path toward the data trees based on the address. The data structure accesses the address in time $O(\log m)$, and each data tree prepares the states in time $O(\log n)$ for the row vector or $O(\log m)$ for the weight vector. The writing process can be completed for the same cost of time. We summarize this data structure in the following definition. 
\begin{definition}\label{access}
    Suppose $A\in\mathbb{R}^{m\times n}$ is a matrix and $\bm \omega\in\mathbb{R}^{2m}$ is a vector. There exists a data structure with the following properties:

    (1) It can be visited by a quantum algorithm that can perform the mapping $V:\ket{i}\ket{0}\rightarrow\ket{i}\ket{A_i}$ for $i\in\{1,2,\cdots,m\}$ in time $\operatorname{poly}(\log(m+n))$. 

    (2) It can be visited by another quantum algorithm that can perform the mapping $G:\ket{0}\rightarrow\ket{\bm \omega}$ in time $\operatorname{poly}(\log m)$.
\end{definition}

\subsection{The sketch of the algorithm}\label{The sketch of the algorithm}

From the above subsections, we already have the way to construct the iteration matrix and the efficient way to prepare the states. It's enough to design the quantum multi-row iteration algorithm. 

The explicit sketch of the implementation of the multi-row iteration algorithm is stated as follows. We will explain the algorithm step by step.
\begin{breakablealgorithm}
    \begin{algorithmic}[1]\label{algorithm1}
    \caption{Quantum multi-row iteration algorithm}
    \label{alg1}
    
        \REQUIRE \textbf{Input:} Randomly choose a unit vector $x^1$. Set $k=1$, $v_1=1$, and the maximum number of iteration steps is $K$.
        
        \ENSURE \textbf{Output:} $\ket{0}^{\otimes (K-1)}\ket{0}^{\otimes\log \lceil q\mod m\rceil}\ket{x^K}+\ket{Gb}$, where $\ket{Gb}$ is the garbage state.
        
        \textbf{Procedure:}
        \STATE Randomly choose $q$ elements from set $\{0,1,\cdots,m-1\}$ as an index set $\tau_1$. The weights $s_{1,i},i\in\tau_1$ are set uniformly. Prepare the state
        \begin{equation}
            \ket{X^1}=\frac{\|x^1\|}{v_1}\sum_{i\in\tau_1}s_{1,i}\ket{i}\ket{x^1}
        \end{equation}
        \STATE Define $\beta_k=\frac{v_k}{\sqrt{v_k^2+\sum_{i\in\tau_k}\|b_i\|^2}}$ and $\gamma_{k,i}=\frac{b_i}{\sqrt{v_k^2+\sum_{i\in\tau_k}\|b_i\|^2}}$, $i\in\tau_k$. Set $v_{k+1}=\frac{v_k}{\beta_{k}}$. Then, through some rotation gates and controlled operation, obtain the state
        \begin{equation}
            \ket{Y^k}=\beta_k\ket{0}\ket{X^k}+\ket{1}\ket{0}^{\otimes (k-1)}\otimes\sum_{i\in\tau_k}\gamma_{k,i}s_{k,i}\ket{i}\ket{A_i}.
        \end{equation}
        \STATE Apply $(I_2^{\otimes (k-1)}\otimes U_k)SWAP_{1,k}$ to $\ket{Y^k}$, then we can obtain
        \begin{equation}
            \ket{Z^{k+1}}=\frac{\|x^{k+1}\|}{v_{k+1}}\ket{0}^{\otimes k}\otimes\sum_{i\in\tau_k}s_{k+1,i}'\ket{i}\ket{x^{k+1}}+\ket{Gb}.
        \end{equation}
        \STATE Randomly choose $q$ elements from set $\{0,1,\cdots,m-1\}$ as an index set $\tau_{k+1}$ with $k=1,2,3,\cdots,K-1$. Implement the exchange operator $P$ to swap the ancillas from $\sum_{i\in\tau_{k}}\ket{i}$ to $\sum_{i\in\tau_{k+1}}\ket{i}$, then obtain the following state
        \begin{equation}
            \ket{X^{k+1}}=\frac{\|x^{k+1}\|}{v_{k+1}}\ket{0}^{\otimes {k}}\otimes\sum_{i\in\tau_{k+1}}s_{{k+1},i}\ket{i}\ket{x^{k+1}}+\ket{Gb}.
        \end{equation}
        \STATE Set $k=k+1$, if the maximum number of iterations is not satisfied, turn to 2; else, turn to 6.
        \STATE Perform an adding procedure $U_{ADD}$, then we obtain the output state.
    \end{algorithmic}
\end{breakablealgorithm}

We prepare the initial state in step 1. In step 2, we define the parameters $\beta_k$, $\gamma_{k,i}$ and $v_k$. $\gamma_{k,i}$ helps to apply $b_i$, $v_k$ is the normalized factor of each iteration and $\beta_k$ is introduced for the purpose of normalization. We can first prepare $\beta_k\ket{0}\sum_{i\in\tau_k}\ket{i}\ket{0}+\ket{1}\sum_{i\in\tau_k}\gamma_{k,i}\ket{i}\ket{0}$ through some rotation gates, then use the controlled operation to apply $\ket{X^k}$ and $\ket{A_i}$.

In step 3, we can obtain the following state after applying the SWAP gate
\begin{widetext}
    \begin{small}
\begin{equation}
\begin{aligned}\label{swapu}
    &SWAP_{1,k} \left(\beta_k\ket{0}\ket{X^k}+\ket{1}\ket{0}^{\otimes (k-1)}\otimes\sum_{i\in\tau_k}\gamma_{k,i}s_{k,i}\ket{i}\ket{A_i}\right)\\
    =&\ket{0}^{\otimes (k-1)} \left(\|\bm x^k\|\beta_k\ket{0}\sum_{i\in\tau_k}s_{k,i}\ket{i}\ket{x^k}+\ket{1}\sum_{i\in\tau_k}\gamma_{k,i}s_{k,i}\ket{i}\ket{A_i}\right)
\end{aligned}
\end{equation}
\end{small}
Then, applying the operator $U_k$ yields
\begin{equation}
\begin{aligned}\label{appU}
&(I_2^{\otimes (k-1)}\otimes U_k) \left(\|\bm x^k\|\beta_k\ket{0}\sum_{i\in\tau_k}s_{k,i}\ket{i}\ket{x^k}+\ket{1}\sum_{i\in\tau_k}\gamma_{k,i}s_{k,i}\ket{i}\ket{A_i}\right)\\
=&\ket{0}^{\otimes k} \Big{(}\|\bm x^k\|\beta_k(I-\Sigma_{\tau_k})\sum_{i\in \tau_k}s_{k,i}\ket{i}\ket{x^k}+\sum_{i\in\tau_k}\gamma_{k,i}s_{k,i}\ket{i}\ket{A_i}\Big{)}+\ket{Gb}\\
=&\frac{\beta_k}{v_k}\ket{0}^{\otimes k}\Big{(}\|\bm x^k\|(I-\Sigma_{\tau_k})\sum_{i\in \tau_k}s_{k,i}\ket{i}\ket{x^k}+\sum_{i\in\tau_k}\omega_{k,i}b_{i}s_{k,i}\ket{i}\ket{A_i}\Big{)}+\ket{Gb}\\
=&\ket{Z^{k+1}}
\end{aligned}
\end{equation}
\end{widetext}
The first term of the third line is quite similar to the result given in Eq.(\ref{rerererer}) with a slight difference in the normalization factor.
Then, in step 4, we can move the index register to a new subspace with an exchange operator $P$
\begin{equation}
P\sum_{i\in\tau_k}s_{k+1,i}'\ket{i}\ket{x^{k+1}}=\sum_{i\in\tau_{k+1}}s_{k+1,i}\ket{i}\ket{x^{k+1}}
\end{equation}
The above process completes the iteration of a step.
Finally, once a predetermined number of iterations or other termination criteria have been met, a quantum-adding procedure is carried out to derive the ultimate outcome
\begin{equation}
    U_{ADD}\sum_{i\in\tau_k}s_{k,i}\ket{i}\ket{x^k}=\ket{0}\ket{x^k}.
\end{equation}
Such an operator can be applied; for example, we can apply a set of Hadamard gates to accomplish this with the help of oblivious amplitude amplification~\cite{berry2015hamiltonian} or design an exact operator to achieve this.

\begin{theorem}\label{dinglili}
    Assume the memory access operator $G$ and $V$ as defined in Definition.\ref{access}. In Algorithm.\ref{algorithm1}, for any $K\ge 1$, the time complexity to prepare $\ket{X^K}$ is
    \begin{equation}
        O(K\log m).
    \end{equation}
\end{theorem}

\section{Analysis for the resource requirement}\label{anan}

\setlength{\parskip}{0.2cm plus4mm minus3mm}
In the preceding section, we give the sketch of the algorithm and the main techniques we use to design the explicit circuit. In this section, we analyse the resource requirement of the algorithm. We suppose the summation of the weights satisfy $\sum_{i\in\tau_k}\omega_{k,i}=t_k$ and the matrix $A$ is given as $A\in\mathbb{R}^{m\times n}$. The outline of this section is as follows. In Sec.\ref{entire}, we analyze the resources needed for $k-th$ iteration step. Then, we analyze the iteration steps needed in Sec.\ref{stepss}, which are almost the same as the classic scenario.

\subsection{Analysis for $k-th$ iteration step}\label{entire}

\begin{figure}[h!]
    \centering
    \includegraphics[width=\linewidth]{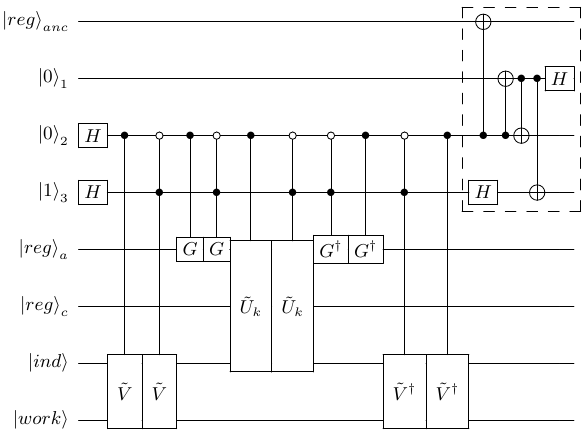}
    \caption{Quantum circuit implementation of the operator $U_k$.}
    \label{IMU}
\end{figure}

The techniques which have been shown before can fulfill the box in the following equation,
\begin{equation}\label{leg}
\begin{aligned}
    U_k=&I_2\otimes\big(I-\boxed{\tilde{V}\sum_{i\in\tau_k}\omega_{k,i}\ket{i}\ket{0}\bra{0}\bra{i}\tilde{V}^\dagger}\big)\\+&X\otimes\big(\boxed{\tilde{V}\sum_{i\in\tau_k}\omega_{k,i}\ket{i}\ket{0}\bra{0}\bra{i}\tilde{V}^\dagger}\big)
\end{aligned}
\end{equation}
To implement the operator $U_k$, we should implement a controlled version of $\tilde{U}_k$ and controlled memory access operator $G$ and $\tilde{V}$. The circuit for $U_k$ is shown in Fig.\ref{IMU}. 
The circuit before the dotted box in Fig.\ref{IMU} applies the operator in the box of Eq.(\ref{leg}) on the states marked by $\ket{010},\ket{011},\ket{001}$.
Then, applying the circuit in the dotted box, we realize the operator $U_k$. More details are given in Appendix.\ref{UEQ}.

\begin{remark}
    This circuit may trigger confusion for some readers because it doesn't look as symmetrical as some of the common quantum circuits. However, because the quantum comparators have a symmetric structure, it's quite possible that we can design a symmetrical structure and set the middle part as a multi-qubit-controlled NOT gate, which completes $I_2\otimes(I-\sum_{i\in\tau_k}\ket{i}\bra{i})+X\otimes\sum_{i\in\tau_k}\ket{i}\bra{i}$. But, this involves the specific design of multiple quantum comparators combined, therefore we leave this to future work.
\end{remark}

The complexity to apply the classical multi-row iteration algorithm for one step is $O(m)$~\cite{moorman2021randomized}.
For the quantum multi-row iteration algorithm, as shown in Algorithm.\ref{algorithm1}, the complexity to complete one iteration step is $O(\log m)$. This exhibits an exponential speedup. For a more formal version, the complexity of the quantum multi-row iteration algorithm for one iteration step is given by the following theorem.
\begin{theorem}
    Given a system of linear equations, $Ax=b,A\in\mathbb{R}^{m\times n}$, the operator $G$ and $\tilde{V}$. For any $k-th$ step, the quantum multi-row iteration algorithm can output the state $\ket{x^{k+1}}$ with high probability using $O(\sqrt{\frac{V_{k+1}^2}{t_{k}}})$ queries to $G$ and $O(\sqrt{V_{k+1}^2})$ queries to $\tilde{V}$, $O(\sqrt{\frac{V_{k+1}^2}{t_k}}\log m)$ extra elementary gates and $O(\log m)$ ancillary qubits.
\end{theorem}
The proof of the theorem is given in Appendix.\ref{Details of the whole process}

\subsection{Analysis for the number of iteration steps}\label{stepss}

Sec.\ref{class_one} provides the condition, $PWD^{-2}=\frac{\alpha_A}{\|A\|_F^2}I$ , to choose the proper selection probability, which affects the address tree, and the iteration weights. Here we analyse the difference in the convergence rate between the classical setting and the quantum setting. 

Without loss of generality, we assume that the norm of each row of the matrix $A$ satisfies $\|A_i\|=1$. This gives the condition $PW=\frac{\alpha_A}{m}I$. In the quantum setting, we have the condition $\sum_{i\in\tau_k}\omega_{k,i}\le1$, because of the need for normalization. As shown in Sec.\ref{Overcome the non-unitarity}, $\omega_{k,i}$ is a modified term, which is determined by $\omega_i$ and $q$. Therefore, in the quantum setting, the choice of weights should satisfy $\sum_{i\in\tau_k}\omega_i\le q$. If any row is selected with equal probability and each selected row has the same iteration weight $\omega_i\le1$, then we have the condition $PW\le\frac{1}{m}I$. It should be noted that $\omega_i\le1$ is a condition derived from $\sum_{i\in\tau_k}\omega_i\le q$, since $\tau_k$ has $q$ elements (may have duplicate elements). Therefore, the parameter $\alpha_A$ should satisfy $\alpha_A\le1$. This is the difference between the classical setting and the quantum setting. We can obtain the convergence rate in the quantum setting based on this.
\begin{lemma}[Convergence rate in the quantum setting]
\label{lemmaquantum}
 Given the quantum multi-row iteration algorithm and suppose the matrix $P$ and $W$ in Definition.\ref{deff} are chosen such that $PWD^{-2}=\frac{\alpha_A}{\|A\|_F^2}I$, the convergence rate of the algorithm satisfies
    \begin{equation}
    \begin{aligned}
        \mathbb{E}&[\|\bm e^{k+1}\|^2]\\&\le\sigma_{max}\left((I-\alpha_A\frac{A^TA}{\|A\|_F^2})^2-\frac{\alpha_A^2}{q}(\frac{A^TA}{\|A\|_F^2})^2\right)\|\bm e^k\|^2\\&+\frac{\alpha_A}{q}\frac{\|\bm r^k\|_W^2}{\|A\|_F^2} 
    \end{aligned}
\end{equation}
where $\bm e^k=\bm x^k-\bm x^*$ is the iteration error between $\bm x^k$ and the solution or least-square solution $\bm x^*$ of step $k$, $\sigma_{max}$ is the maximum singular value, $\|A\|_F^2=\sum_{i,j}A_{ij}^2$, $0<\alpha_A\le 1$ is the relaxation parameter, $\|\cdot\|_W^2=\braket{\cdot,W}$, and $\bm r^k:=\bm b-A\bm x^k$ is the residual of the $k$th iteration. 
\end{lemma}
The proof of the convergence rate is given in Appendix.\ref{Proof of the convergence rate}. The expression for the convergence rate is a little bit complex and therefore it's hard to tell how much the quantum multi-row iteration algorithm outperforms the quantum one-row iteration algorithm. We will show this in the next section by numerical test. Also, the influence of the difference of the parameter $\alpha_A$ will also be shown numerically.

\section{Numerical results}\label{Numerical experiment}

In this section, we conduct some numerical tests to show the advantage of the quantum multi-row iteration over the quantum one-row iteration. Moreover, we show the effect of the parameter $\alpha_A$ to illustrate the difference between the quantum algorithm and the classic one.

For each numerical test, we run 100 independent trials and evaluate the average squared error norms $\|\bm e^k\|^2$ across the trials, where $\bm e^k=\bm x^k-\bm x^*$. We get the iteration result $\bm x^k$ by multiplying the value $v_{k+1}$ with the state vector $\ket{\bm x^k}$, which is obtained through the $\textbf{state-vector}$ compiler in $\textbf{qiskit}$. The matrix $A$ is a $100\times 4$ Gaussian matrix with each row normalized. The least-squares solution $\bm x^*$ is a 4-dimensional Gaussian vector and satisfies $\|\bm x^*\|=1$. We randomly choose a residual $\bm r^*$ and normalize it so that $\|\bm r^*\|=1$. $\bm b$ is computed as $\bm r^*+A\bm x^*$.

\begin{figure}[h!]
    \centering
    \includegraphics[width=\linewidth]{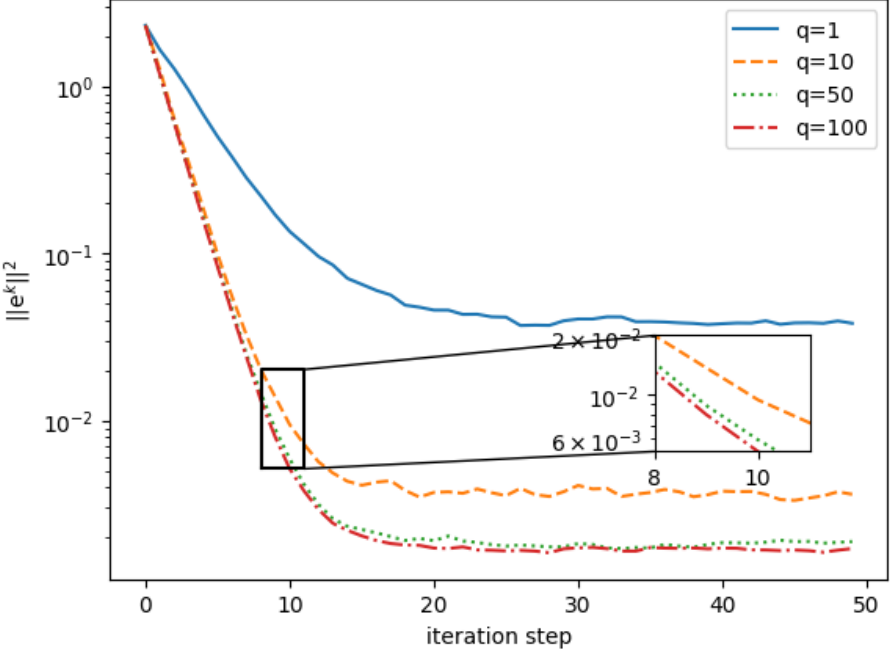}
    \caption{Comparison of the quantum one-row iteration and the quantum multi-row iteration with different choices of the number of iteration rows. $q$ is the number of rows selected in each iteration. $e^k=x^k-x^*$ is the error between the iteration result and the solution or least-square solution. Smaller $\|e^k\|$ indicates a smaller convergence radius.}
    \label{outputtt}
\end{figure}

In Fig.\ref{outputtt}, we compare the quantum one-row iteration with the quantum multi-row iteration, which uses uniform weights for simplicity. The figure shows that the quantum multi-row iteration possesses a better convergence rate and a smaller convergence radius. Moreover, the quantum multi-row iteration can reach a better convergence rate as more rows are selected.

\begin{figure}
    \centering
    \includegraphics[width=\linewidth]{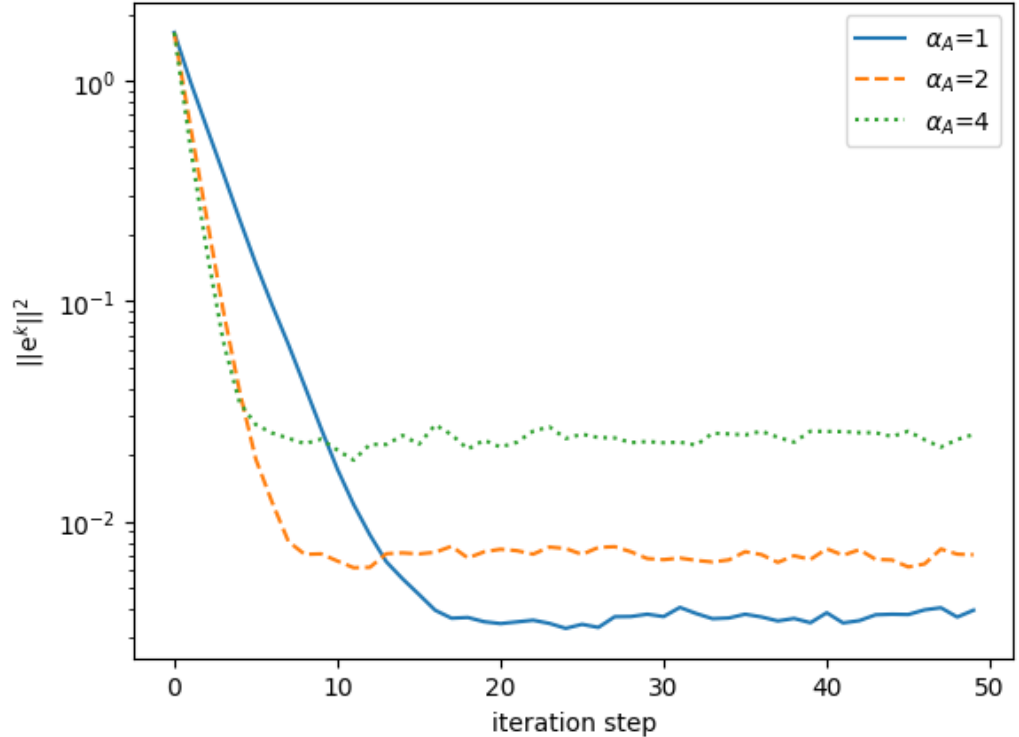}
    \caption{Comparison of the effect of different choices of $\alpha_A$. The classic algorithm achieves the results with $\alpha_A>1$. We choose uniform weights, and the number of iteration rows satisfies $q=10$. $e^k=x^k-x^*$ is the error between the iteration result and the solution or least-square solution.}
    \label{outputt}
\end{figure}

In Fig.\ref{outputt}, we compare the effect of different choices of $\alpha_A$. The classic algorithm achieves the results with $\alpha_A>1$. When $\alpha_A$ reaches a critical point, the algorithm can have a slightly better convergence rate. The larger choice of $\alpha_A$ has a worse convergence radius. It's shown in~\cite{moorman2021randomized} that the optimal choice of $\alpha_A$ is not large. Therefore, the drawback that the quantum multi-row iteration cannot choose a large $\alpha_A$ is not severe.

\section{More applications}\label{Application}

Our method not only serves as a solver for overdetermined equations but also functions as a solver for linear systems and stochastic gradient descent.

If $Ax=b,A\in\mathbb{R}^{n\times n}$, our methods can serve as a linear system solver. There is no need to assume that the equations are consistent, which implies that at least one solution to the given equation exists because the randomized Kaczmarz iteration method has the potential to solve the inconsistent problem.

Using the algorithm defined in the previous section, we give the following theorem.
\begin{corollary}
    Given a system of linear equations, $Ax=b,A\in\mathbb{R}^{n\times n}$. Suppose the iteration error $\|\bm e^k\|\le\epsilon$ is satisfied after $K$ iteration steps. The quantum algorithm defined in Algorithm.\ref{alg1} can prepare the state that encodes the solution or least square solution of such equations within the error $\|e^K\|$ with high probability. The query complexity is $O(K)$ and the gate complexity is $O(K\log n)$. The iteration step $K$ has a maximum value $K_{max}=\kappa_s^2\log(\frac{1}{\epsilon})$ when $q=1$, where $\kappa_s=\|A\|_F\|A^{-1}\|$ and $\|A\|_F$ is the Frobenius norm.
\end{corollary}

The term $\kappa_s=\|A\|_F\|A^{-1}\|$, where $\|A\|_F$ is the frobenius norm of the matrix $A$ and $\epsilon$ is the tolerance error. $k_{max}$ only occurs when only one selected iteration row ($q=1$) exists.

The randomized Kaczmarz iteration can be viewed as a subcase of stochastic gradient descent for the following loss function~\cite{needell2014stochastic}
\begin{equation}\label{loss_fun}
    F(x)=\sum^n_{i=1}f_i(x)=\sum^n_{i=1}\frac{1}{2}(a_ix-b_i)^2
\end{equation}
which covers the case when the gradient is an affine function for quadratic optimization problems of the form $min_{x\in\mathbb{R}^n}x^TAx+b^Tx+c$ for $A\in\mathbb{R}^{n\times n},b\in\mathbb{R}^n,c\in\mathbb{R}$~\cite{kerenidis2020quantum}.
Therefore, the randomized kaczmarz iteration method is a reweighted version of the stochastic gradient descent
\begin{equation}
\begin{aligned}
    x^{k+1}&=x^k-\frac{a_ix^k-b_i}{\|a_i\|^2}a_i^T\\
    &=x^k-\frac{\bigtriangledown f_i(x)}{\|a_i\|^2}
\end{aligned}
\end{equation}
The multi-row iteration method can be seen as mini-batch stochastic gradient descent~\cite{moorman2021randomized}
\begin{equation}
\begin{aligned}
    x^{k+1}&=x^k-\frac{1}{q}\sum_{i\in\tau_k}\omega_i\frac{a_ix^k-b_i}{\|a_i\|^2}a_i^T\\
    &=x^k-\frac{1}{q}\sum_{i\in\tau_k}\frac{\omega_i}{\|a_i\|^2}\bigtriangledown f_i(x)
\end{aligned}
\end{equation}
Thus, we summarize this idea as the following theorem.
\begin{corollary}
    Given a loss function as defined in (\ref{loss_fun}), the quantum algorithm described in Algorithm.\ref{alg1} can prepare a quantum state that encodes the result of $K$ iterations of stochastic gradient descent of the loss function with query complexity $O(K)$ and gate complexity $O(K\log n)$.
\end{corollary}

\section{Summary}
In this paper, we present a quantum algorithm for a linear system with non-square coefficient matrix. We show that the quantum version of multi-row iterations possesses exponential speedups in problem size $n$ and a faster convergence rate in the constraints $m$ while keeping the logarithmic dependence on the error tolerance. There are still many open questions. For example, combining the row and column iteration shows a faster convergence rate. Still, the pure quantum version of this method does not exist because this poses new challenges for both state generation and circuit implementation. Is there a proper quantum data structure for this method? Moreover, for many iteration methods with different strategies, such as iteration with a small block, can we construct pure quantum version algorithms for these methods? Moreover, is there a fast and general quantum algorithm for all kinds of linear systems? What is the lower bound on the complexity of such an approach? 

\begin{acknowledgments}
This work was supported in part by the National Natural Science Foundation of China Grants No. 62325210, 62272441, 12204489, 62301531, and the Strategic Priority Research Program of Chinese Academy of Sciences Grant No.XDB28000000.
\end{acknowledgments}

\appendix
\begin{appendices}
\section{Details of the application of $U_k$}\label{UEQ}

As shown in Sec.\ref{Equivalent implementation}, applying the iteration matrix $U_k$ requires to apply the block-encoding given in Eq.(\ref{ttototo}). Such block-encoding achieves the linear combination of unitary
\begin{equation}
    \sum_{i\in\tau_k}(\frac{\omega_{k,i}}{2}C_1^{(i,i)}-\frac{\omega_{k,i}}{2}C_{-1}^{(i,i)})+\sum_{i\notin\tau_k}(\frac{r_{k,i}}{2}C_1^{(i,i)}-\frac{r_{k,i}}{2}C_{1}^{(i,i)})
\end{equation}
We introduce an ancilla register to help apply the coefficients. Then, the combination can be treated as
\begin{equation}
\begin{aligned}
    \tilde{U}_k=&\sum^{m-1}_{j=0}\ket{j}\bra{j}\otimes C^{(j,j)}_1+\sum^{2m-1}_{\substack{j=m\\j-m\in\tau_k}}\ket{j}\bra{j}\otimes C^{(j-m,j-m)}_{-1}\\
    +&\sum^{2m-1}_{\substack{j=m\\j-m\notin\tau_k}}\ket{j}\bra{j}\otimes C^{(j-m,j-m)}_{1}
\end{aligned}
\end{equation}
with the coefficients applying on different indices $j$.
We can apply the operator $\tilde{U}_k$ equivalently by an operator $U_{eq}$, as they achieve the same result on the basis state
\begin{widetext}  
        \begin{equation}\label{uuuuuk}
        \tilde{U}_k\ket{j}\ket{l}=\left\{
        \begin{array}{ll}
        \ket{j}\ket{(2j-l)\mod m},&0\le j\le m-1\\
        \ket{j}\ket{(2j-l)\mod m},&m\le j\le 2m-1\\
        -\ket{j}\ket{(2j-l)\mod m},&m\le j\le 2m-1,l=(j\mod m),j-m\in\tau_k
        \end{array}
        \right.
    \end{equation}
\end{widetext}

We define the following function
\begin{equation}
        f(j,l)=\left\{
        \begin{array}{cc}
            0,&0\le j\le m-1\\
            0,&m\le j\le 2m-1\\
            1,&m\le j\le 2m-1,\\&l=(j\mod m),(j\mod m)\in\tau_k
        \end{array}
        \right.
    \end{equation}
This function is associated with the flip on the corresponding state. To compute such classical function with a quantum circuit, we use the quantum comparator~\cite{cuccaro2004new,li2020efficient}. The comparator compares the natural numbers $a$ and $b$ in two registers and outputs the result $c$ in the third register, if $a\le b$, $c=0$; otherwise, $c=1$. The specific circuit of $U_f$ is given as follows.

(1)For $(j\mod m)\in\tau_k$, prepare the initial state
\begin{equation*}
\begin{aligned}  &\ket{j}_{a_1}\ket{l}_{a_2}\ket{0}_{b_1}\ket{0}_{b_2}\ket{0}_{b_3}\ket{0}_{c_1}\ket{0}_{c_2}\ket{0}_{c_3}\ket{-}_{c_4}\\
\rightarrow&\ket{j}_{a_1}\ket{l}_{a_2}\ket{m-1}_{b_1}\ket{(j-1)\mod m}_{b_2}\ket{j\mod m}_{b_3}\\&\ket{0}_{c_1}\ket{0}_{c_2}\ket{0}_{c_3}\ket{-}_{c_4}
\end{aligned}
\end{equation*}

For $(j\mod m)\notin\tau_k$, prepare the initial state
\begin{equation*}
\begin{aligned}  &\ket{j}_{a_1}\ket{l}_{a_2}\ket{0}_{b_1}\ket{0}_{b_2}\ket{0}_{b_3}\ket{0}_{c_1}\ket{0}_{c_2}\ket{0}_{c_3}\ket{-}_{c_4}\\
\rightarrow&\ket{j}_{a_1}\ket{l}_{a_2}\ket{2m-1}_{b_1}\ket{(j-1)\mod m}_{b_2}\ket{j\mod m}_{b_3}\\&\ket{0}_{c_1}\ket{0}_{c_2}\ket{0}_{c_3}\ket{-}_{c_4}
\end{aligned}
\end{equation*}

(2)Perform quantum comparator on registers $\{a_1,b_1,c_1\}$, $\{a_2,b_2,c_2\}$ and $\{a_3,b_3,c_3\}$ respectively.

(3)Perform Toffoli gate on the registers on $\{c_1,c_2,c_4\}$, $\{c_1,c_3,c_4\}$ and $\{c_2,c_3,c_4\}$.

(4)Reverse the computation on registers $\{c_3,c_2,c_1,b_3,b_2,b_1\}$.

The states of registers $c_1,c_2$ and $c_3$ are set as $\ket{1}$, when $a_1>b_1$, $a_2>b_2$ and $a_2\le b_3$ are satisfied respectively.
The mapping on registers $a_1$, $a_2$ and $c_4$ is
\begin{equation}
    U_f\ket{j}\ket{l}\ket{-}=(-1)^{f(j,l)}\ket{j}\ket{l}\ket{-}
\end{equation}
Using a quantum modular adder, we can implement 
\begin{equation}
    U_{adder}\ket{j}\ket{l}=\ket{j}\ket{(2j-l)\mod m}
\end{equation}
Therefore, the equivalent process $U_{eq}$ can be implemented by $U_f$ and $U_{adder}$. Then, the box in Eq.(\ref{leg}) is completed by
\begin{equation}
    (I\otimes\tilde{V})(G\otimes I)\tilde{U}_k(G^\dagger\otimes I)(I\otimes\tilde{V}^\dagger)
\end{equation}
where we omit the subscript of $I$. If we attempt to treat the operator $G$ and $\tilde{V}$ as state preparation operators instead of memory access operators, we should have the following observation.
\begin{observation}\label{nota1}
    In practice, the memory access operator $V$ can be replaced by the state preparation operator. However, besides satisfying the assumption $V\ket{i}\ket{0}=\ket{i}\ket{A_i}$, the matrix associated with the state preparation operator should possess symmetry.
\end{observation}
The proof is given in Appendix.\ref{PNota1}. 

However, this is not enough for the implementation of the operator $U_k$. We need to implement a controlled version of $\tilde{U}_k$ and the memory access operator $G$ and $\tilde{V}$. Therefore, we introduce two extra qubits, and then we can prepare the following state
\begin{equation}\label{targ}
\begin{aligned}
    &\ket{00}_{anc}\ket{ind}\ket{work}\\-&\ket{01}_{anc}\tilde{V}\left(\braket{0|(G\otimes I)\tilde{U}_k(G^\dagger\otimes I)|0}\right)\tilde{V}^\dagger\ket{ind}\ket{work}\\
    +&\ket{10}_{anc}\tilde{V}\left(\braket{0|(G\otimes I)\tilde{U}_k(G^\dagger\otimes I)|0}\right)\tilde{V}^\dagger\ket{ind}\ket{work}\\-&\ket{11}_{anc}\tilde{V}\left(\braket{0|(G\otimes I)\tilde{U}_k(G^\dagger\otimes I)|0}\right)\tilde{V}^\dagger\ket{ind}\ket{work}.
\end{aligned}
\end{equation}
Introducing one more ancillary qubit and applying some Hadamard gates and CNOT gates, we can obtain the required states after applying the operator $U_k$. The circuit for the above state is shown in Fig.\ref{IMU}. This equals to a (1,3,$\epsilon$)-block-encoding of the operator $U_k$.
The circuit before the dotted box in Fig.\ref{IMU} prepares the state in (\ref{targ}). To simplify the notation, we use $\ket{target}$ to represent the state $\tilde{V}(\braket{0|G\tilde{U}_kG^\dagger|0})\tilde{V}^\dagger\ket{ind}\ket{work}$ and omit the register $a$ and $c$. Then, the circuit in the dotted box fulfills the following mapping
\begin{widetext}
\begin{equation}
\begin{aligned}
    &\ket{anc}\ket{0}(\ket{00}\ket{ind}\ket{work}-\ket{01}\ket{target}+\ket{10}\ket{target}-\ket{11}\ket{target})\\
    \rightarrow &\left(I_2\otimes\big(I- \tilde{V}(\braket{0|(G\otimes I)\tilde{U}_k(G^\dagger\otimes I)|0})\tilde{V}^\dagger\big)+X\otimes \tilde{V}\big(\braket{0|(G\otimes I)\tilde{U}_k(G^\dagger\otimes I)|0}\big)\tilde{V}^\dagger\right)\\&\ket{anc}\ket{000}\ket{ind}\ket{work}+\ket{Gb}.
\end{aligned}
\end{equation}
\end{widetext}
Therefore, we fulfill the construction of the block-encoding of operator $U_k$.

It should also be noted that if the weights satisfy $\sum_{i\in\tau_k}\omega_{k,i}=1$. We suppose the operator $G$ completes $G\ket{0}=\sum_{i\in\tau_k}\omega_{k,i}\ket{i}$, then apply the operator $G$ and $\tilde{V}$, a multi-qubit-controlled NOT gate, which completes $I_2\otimes(I-\sum_{i\in\tau_k}\ket{i}\bra{i})+X\otimes\sum_{i\in\tau_k}\ket{i}\bra{i}$ and $G^\dagger$ and $\tilde{V}^\dagger$. This can achieve the same result.

\section{Details of the whole process}\label{Details of the whole process}

\subsection{Implementation of the whole process}

\begin{figure}[h!]
    \centering
    \includegraphics[width=\linewidth]{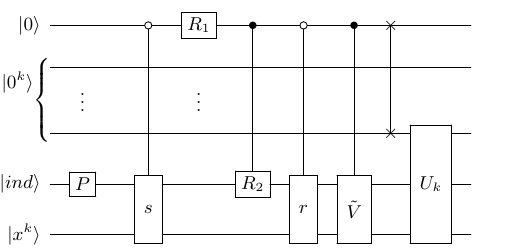}
    \caption{Implementation of an iteration step with the help of QRAM, where $s$ represents the process for storage and $r$ for the reading process. The operators $R_1$ and $R_2$ are the rotation operators.}
    \label{oneiter1}
\end{figure}

Fig.\ref{oneiter1} shows the circuit of an iteration step. It requires a storage process $s$ and a reading process $r$. That's because at the end of any iteration step $k$, we obtain the state $\ket{X^{k+1}}$. However, at the beginning of any step $k$, we are required to prepare $\beta_k\ket{0}\sum_{i\in\tau_k}\ket{i}\ket{0}+\ket{1}\sum_{i\in\tau_k}\gamma_{k,i}\ket{i}\ket{0}$ and apply $\ket{X^k}$ and $\ket{A_i}$. Therefore, we should store the result from the last iteration step. This can be avoided if we apply the rotation at the beginning of the whole process.

\textit{1. prepare at each step.}-
To generate the state $\ket{Y^k}$, we can first prepare $\beta_k\ket{0}\sum_{i\in\tau_k}s_{k,i}\ket{i}\ket{0}+\ket{1}\sum_{i\in\tau_k}\gamma_{k,i}s_{k,i}\ket{i}\ket{0}$ using a set of rotation operators. Then, perform a controlled operator to obtain $\ket{X^k}$ and $\ket{i}\ket{A_i}$ on the index and work register with the help of QRAM(see Fig.\ref{oneiter1}). At the end of each iteration, we need to store the current result, reset the working register, and then repeat the above process. This would require extra access to the QRAM compared to the other idea, as we need to store the result after each iteration. 

\textit{2. prepare before the iteration.}-
The other idea is to prepare $\theta_1\ket{0\cdots 0}\sum_{i\in\tau_1}\ket{i}\ket{x^1}+\theta_2\ket{0\cdots 01}\sum_{i\in\tau_1}\ket{i}\ket{A_i}+\theta_3\ket{0\cdots 010}\sum_{i\in\tau_2}\ket{i}\ket{A_i}+\cdots$ before starting the iteration procedure. The required rotation gates for $\theta$, which depends on the choice of weights, and each set $\tau_k$ that depends on the choice of rows, can be obtained through a pre-processing procedure. This will require that the memory access procedure become multi-qubit-controlled and the iteration operator become one-qubit-controlled.

\subsection{Resource analysis for the whole process}

In the algorithm given in Algorithm.\ref{algorithm1}, the number of quantum gates used consists of four components: the rotation gates for $\beta_k$ and $\gamma_{k,i}$, gates for the iteration matrix $U_k$, gates for the operator $P$ and gates for $U_{ADD}$. 

For each step, the rotation requires $O(\log m)$ elementary gates. The application of the quantum comparators and the quantum modular adder needs $O(\log m)$ elementary gates and $O(\log m)$ ancilla qubits. The operator $\braket{0|(G\otimes I)\tilde{U}_k(G^\dagger\otimes I)|0}$ can be applied with a probability of $\sum_{i\in\tau_k}\omega_{k,i}=t_k$, then we can apply amplitude amplification $O(\frac{1}{\sqrt{t_k}})$ times to achieve a high probability. Therefore, the iteration matrix $U_k$ needs $O(\frac{1}{\sqrt{t_k}})$ queries to $G$ and $O(1)$ queries to $\tilde{V}$, $O(\frac{1}{\sqrt{t_k}}\log m)$ elementary gates and $O(\log m)$ ancilla qubits to complete. Through appropriate selection of weights $\sum_{i\in\tau_k}\omega_{k,i}=t_k$, $\frac{1}{\sqrt{t_k}}$ is a constant and can be neglected. We omit the resource for the operator $P$, since it depends on specific situations, but in most cases, several SWAP gates and controlled NOT gates are enough. 

We also omit the resource for the operator $U_{ADD}$, since it also depends on specific situations. If we use $\log m$ Hadamard gates to apply $U_{ADD}$, then we will require to apply amplitude amplification $O(\sqrt{m/\sum_{i\in\tau_{K}}s_{K,i}})$ times, where $K$ is the number of the total iteration steps, to achieve the result with a high probability. But, if we can design an exact operator that achieves the same effect, then the resource required will be quite different. 

For an iteration process with $K$ iteration steps, we can obtain the final result with a probability of $\frac{1}{V_K^2}$. In order to obtain the result with a high probability, we need to perform amplitude amplification $O(\sqrt{V_K^2})$ times. It should be noted that $V_K=\sqrt{1+\sum_{k=1}^K\sum_{i\in\tau_k}b_i}$ is not large, because the parameter $b_i$ here is rescaled as mentioned in Sec.\ref{Overcome the non-unitarity} and the number of iteration steps is limited.
In summary, $K$ iteration steps require $O(K\sqrt{\frac{V_K^2}{t_K}})$ queries to $G$ and $O(K\sqrt{V_K^2})$ queries to $\tilde{V}$, $O(K\sqrt{\frac{V_K^2}{t_K}}\log m)$ elementary gates and $O(\log m+K)$ ancilla qubits (each iteration requires one extra ancilla qubit). Combining with the complexity shown in Definition.\ref{access}, the complexity of the quantum multi-row iteration algorithm is $O(K\sqrt{\frac{V_K^2}{t_K}}\log m)$.

\section{Proof of the convergence rate}\label{Proof of the convergence rate}

The analysis for the convergence rate in the quantum setting is quite similar to the one in the classical setting~\cite{moorman2021randomized}.
To analyse the convergence rate, we begin with the error update at each iteration. The error is defined as $e^k=x^k-x^*$, where $x^*$ is the solution or least-square solution. We suppose the residual is $r^*$ and $Ax^*+r^*=b$. Then, using $A_ie^k-r^*_i=A_ix^k-b_i$, we arrive the error update
\begin{equation}
    e^{k+1}=e^k-\sum_{i\in\tau_k}\omega_{k,i}\frac{A_ie^k-r^*_i}{\|A_i\|^2}A_i^T
\end{equation}
where we suppose $\|A_i\|=1$. We don't omit it for the completeness of the proof.

Define the weighted sampling matrix
\begin{equation}
    M_k:=\sum_{i\in\tau_k}\omega_{k,i}\frac{I_i^TI_i}{\|A_i\|^2}
\end{equation}
Then, the error update can be rewritten as
\begin{equation}
    e^{k+1}=(I-A^TM_kA)e^k+A^TM_kr^*
\end{equation}
To evaluate the error update, we give the following lemma.
\begin{lemma}[\cite{moorman2021randomized}]\label{hahahalemma}
    Given $D$, $P$ and $W$ as defined in definition.\ref{deff}, then we have
    \begin{equation}
        \mathbb{E}[M_k]=PDW^{-2}
    \end{equation}
    and
    \begin{equation}
        \mathbb{E}[M_k^TAA^TM_k]=\frac{1}{q}PW^2D^{-2}+(1-\frac{1}{q})PWD^{-2}AA^TPWD^{-2}
    \end{equation}
\end{lemma}
The proof for this lemma is given in Appendix.\ref{hahaha}. Since $M_k$ is a sample average, as the number of samples goes to infinity, we should have 
\begin{equation}
    M_k\rightarrow PWD^{-2}
\end{equation}
Therefore, as the number of samples goes to infinity, the error update approaches the deterministic update
\begin{equation}
    e^{k+1}=(I-A^TPWD^{-2}A)e^k+A^TPWD^{-2}r^*
\end{equation}
Since we want the error to converge to zero, we should require that this limiting error update has the zero vector, which is
\begin{equation}
    A^TPWD^{-2}r^*=0
\end{equation}
for any least-squares residual $r^*$. This holds when the following equation is satisfied
\begin{equation}
    PWD^{-2}=\alpha_A I
\end{equation}
for $0<\alpha_A\le 1$ ($\alpha_A>0$ is the same as the classical setting and $\alpha_A\le1$ is unique in the quantum setting). The squared error norm is
\begin{equation}
\begin{aligned}
    \|e&^{k+1}\|^2=\|(I-A^TM_kA)e^k+A^TM_kr^*\|^2\\
    =&\|(I-A^TM_kA)e^k\|^2+2\langle (I-A^TM_kA)e^k,A^TM_kr^*\rangle\\
    +&\|A^TM_kr^*\|^2
\end{aligned}
\end{equation}
Taking expectations, we can get
\begin{equation}
\begin{aligned}
    \mathbb{E}[\|e^{k+1}\|^2]=&\mathbb{E}[\|(I-A^TM_kA)e^k\|^2]+\mathbb{E}[\|A^TM_kr^*\|^2]\\
    +&2\mathbb{E}[\langle (I-A^TM_kA)e^k,A^TM_kr^*\rangle]
\end{aligned}
\end{equation}
Using lemma.\ref{hahahalemma}, we can simplify the first term
\begin{widetext}
\begin{equation}
\begin{aligned}
    \mathbb{E}[\|&(I-A^TM_kA)e^k\|^2]\\
    =&\mathbb{E}[\langle e^k,(I-A^TM_kA)^T(I-A^TM_kA)e^k\rangle]\\
    =&\langle e^k,(I-2A^T\mathbb{E}[M_k]A+A^T\mathbb{E}[M_k^TAA^TM_k]A)e^k\rangle\\
    =&\langle e^k,\left(I-2\alpha_A\frac{A^TA}{\|A\|_F^2}+\frac{\alpha_A}{q}\frac{A^TWA}{\|A\|_F^2}+\alpha_A^2(1-\frac{1}{q})(\frac{A^TA}{\|A\|_F^2})^2\right)e^k\rangle\\
    =&\langle e^k,\left(\Big(I-\alpha_A\frac{A^TA}{\|A\|_F^2}\Big)^2+\frac{A^T}{\|A\|_F^2}\Big(\frac{\alpha_A}{q}W-\frac{\alpha_A^2}{q}\frac{AA^T}{\|A\|_F^2}\Big)\frac{A}{\|A\|_F}\right)e^k\rangle
\end{aligned}
\end{equation}
\end{widetext}
For the second term, we can get
\begin{equation}
    \mathbb{E}[\|A^TM_kr^*\|^2]=\langle r^*,\mathbb{E}[M_k^TAA^TM_k]r^*\rangle=\frac{\alpha_A}{q}\frac{\|r^*\|_W^2}{\|A\|_F^2}
\end{equation}
Similarly, for the third term, we can get
\begin{equation}
    2\mathbb{E}[\langle A^TM_kAe^k,A^TM_kr^*\rangle]=\frac{2\alpha_A}{q\|A\|_F^2}\langle Ae^k,Wr^*\rangle
\end{equation}
Combine three terms together, we have
\begin{equation}
\begin{aligned}
    \mathbb{E}[\|e&^{k+1}\|^2]=\langle e^k,\left(I-\alpha_A\frac{A^TA}{\|A\|_F^2}\right)^2e^k\rangle\\
    +&\langle e^k,\frac{A^T}{\|A\|_F^2}\Big(\frac{\alpha_A}{q}W-\frac{\alpha_A^2}{q}\frac{AA^T}{\|A\|_F^2}\Big)\frac{A}{\|A\|_F}e^k\rangle\\
    -&\frac{2\alpha_A}{q\|A\|_F^2}\langle Ae^k,Wr^*\rangle+\frac{\alpha_A}{q}\frac{\|r^*\|_W^2}{\|A\|_F^2}\\
    =&\langle e^k,\left((I-\alpha_A\frac{A^TA}{\|A\|_F^2})^2-\frac{\alpha_A^2}{q}(\frac{A^TA}{\|A\|_F^2})^2\right)e^k\rangle\\+&\frac{\alpha_A}{q}\frac{\|r^*\|_W^2}{\|A\|_F^2}\\
    \le&\sigma_{max}\left((I-\alpha_A\frac{A^TA}{\|A\|_F^2})^2-\frac{\alpha_A^2}{q}(\frac{A^TA}{\|A\|_F^2})^2\right)\|e^k\|^2\\+&\frac{\alpha_A}{q}\frac{\|r^*\|_W^2}{\|A\|_F^2}
\end{aligned}
\end{equation}
This completes the proof.

\section{Proof of Observation \ref{nota1}}\label{PNota1}

\begin{figure}[h!]
    \centering
    \includegraphics[width=\linewidth]{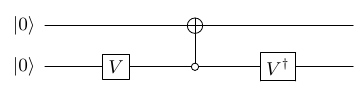}
    \caption{Implementation of the operator $U$.}
    \label{one}
\end{figure}

Without loss of generality, we assume that there exists a state preparation operator $\ket{V}$ satisfying
\begin{equation}
    V\ket{0}=\ket{a}
\end{equation}
where $\ket{a}=a_1\ket{0}+a_2\ket{1}$, without loss of generality, we assume that $a_1$ and $a_2$ are real.
We anticipate utilizing this state preparation operator to achieve
\begin{equation}
\begin{aligned}
    U&=(I_2\otimes V)(I_2\otimes(I-\ket{0}\bra{0})+X\otimes \ket{0}\bra{0})(I_2\otimes V^\dagger)\\
    &=\left[
    \begin{array}{cc}
        I-\ket{a}\bra{a} &\ket{a}\bra{a}  \\
        \ket{a}\bra{a}&I-\ket{a}\bra{a} 
    \end{array}
    \right]
\end{aligned}
\end{equation}
We could represent $(I_2\otimes(I-\ket{0}\bra{0})+X\otimes \ket{0}\bra{0})$ in a matrix format easily,
\begin{equation}
(I_2\otimes(I-\ket{0}\bra{0})+X\otimes \ket{0}\bra{0})=
\left[
\begin{array}{cccc}
    0&0&1&0  \\
    0&1&0&0  \\
    1&0&0&0  \\
    0&0&0&1
\end{array}
\right]
\end{equation}
Then, we can obtain the circuit(Fig.\ref{one}) to perform the operator $U$.
We choose the operator $V$ as
\begin{equation}
V_1=\left[
\begin{array}{cc}
    a_1&a_2  \\
    -a_2&a_1 
\end{array}
\right]
\end{equation}
and
\begin{equation}
V_2=\left[
\begin{array}{cc}
    a_1&a_2  \\
    a_2&-a_1 
\end{array}
\right]
\end{equation}
Both $V_1$ and $V_2$ are unitary operator and satisfy $V\ket{0}=\ket{a}$. By a simple calculation, we obtain
\begin{equation}
\begin{aligned}
    &(I_2\otimes V_1)(I_2\otimes(I-\ket{0}\bra{0})+X\otimes \ket{0}\bra{0})(I_2\otimes V_1^\dagger)\\&=
    \left[
    \begin{array}{cccc}
        -a_2^2&a_1a_2&a_1^2&a_1a_2  \\
        -a_1a_2&a_1^2&-a_1a_2&-a_2^2 \\
        a_1^2&a_1a_2&-a_2^2&a_1a_2\\
        -a_1a_2&-a^2_2&-a_1a_2&a_1^2
    \end{array}
    \right]
\end{aligned}
\end{equation}
and
\begin{equation}
\begin{aligned}
    &(I_2\otimes V_2)(I_2\otimes(I-\ket{0}\bra{0})+X\otimes \ket{0}\bra{0})(I_2\otimes V_2^\dagger)\\&=
    \left[
    \begin{array}{cccc}
        a_2^2&-a_1a_2&a_1^2&a_1a_2  \\
        -a_1a_2&a_1^2&a_1a_2&a_2^2 \\
        a_1^2&a_1a_2&a_2^2&-a_1a_2\\
        a_1a_2&a^2_2&-a_1a_2&a_1^2
    \end{array}
    \right]
\end{aligned}
\end{equation}

Given an input state $\ket{0}\ket{x}$, where $\ket{x}=\ket{0}$(without loss of generality). Applying the operator $U$ on the input state, we will obtain
\begin{equation}
\begin{aligned}
    U\ket{0}\ket{x}&=\ket{0}\ket{x}-\ket{0}\braket{a|x}\ket{a}\\
    &=\ket{0}(1-a_1^2)\ket{0}-\ket{0}a_1a_2\ket{1}\\
    &=\ket{0}a_2^2\ket{0}-\ket{0}a_1a_2\ket{1}
\end{aligned}
\end{equation}
This result can only be obtained if we choose $V=V_2$. And the matrix in (\ref{Uk}) equals the matrix corresponding to the choice of $V_2$.

Extending this statement to the more general case, the definition of $U$ demonstrates the symmetry of $U$. Therefore, in the real number case, the operator $U$ should satisfy $U^T=U$, which indicates that $V=V^\dagger$. This assertion extends to the imaginary number case as $V=(V^\dagger)^*$.

When applying $V\ket{0}\bra{0}V^\dagger$, a similar situation arises. After implementing $(\ket{0}\bra{0})V$ on an arbitrary state $\ket{\psi}$, the resulting state is $\ket{\phi}=(\ket{0}\bra{0})V\ket{\psi}$. This causes the operator $V^\dagger$ to act on the state as $V^\dagger\ket{\phi}$ instead of $\bra{\phi}V^\dagger$. Therefore, it is necessary for the operator $V$ to be symmetric.

\section{Proof of lemma.\ref{hahahalemma} }\label{hahaha}

The expectation of $M_k$ is as follows,
\begin{equation}
\begin{aligned}
    \mathbb{E}[M_k]=&\mathbb{E}[\sum_{i\in\tau_k}\omega_{k,i}\frac{I}{\|A_i\|^2}]=\mathbb{E}[\sum_{i\in\tau_k}\frac{\omega_i}{q}\frac{I}{\|A_i\|^2}]\\
    =&\mathbb{E}[\omega_i\frac{I}{\|A_i\|^2}]=\sum_{i=0}^{m-1}p_i\omega_i\frac{I}{\|A_i\|^2}\\
    =&PWD^{-2}
\end{aligned}
\end{equation}
Similarly, we can compute
\begin{equation}
\begin{aligned}
    \mathbb{E}[&M_k^TAA^TM_k]=\mathbb{E}[(\sum_{i\in\tau_k}\omega_{k,i}\frac{I_i^TA_i}{\|A_i\|^2})(\sum_{j\in\tau_k}\omega_{k,j}\frac{I_j^TA_j}{\|A_j\|^2})]\\
    &=\frac{1}{q}\mathbb{E}[(\omega_i\frac{I_i^TA_i}{\|A_i\|^2})(\omega_i\frac{A_i^TI_i}{\|A_i\|^2})]\\
    &+(1-\frac{1}{q})\mathbb{E}[\omega_i\frac{I_i^TA_i}{\|A_i\|^2}]\mathbb{E}[\omega_j\frac{A_j^TI_j}{\|A_j\|^2}]\\
    &=\frac{1}{q}\mathbb{E}[\omega_i^2\frac{I_i^TI_i}{\|A_i\|^2}]+(1-\frac{1}{q})PWD^{-2}AA^TPWD^{-2}\\
    &=\frac{1}{q}PW^2D^{-2}+(1-\frac{1}{q})PWD^{-2}AA^TPWD^{-2}
\end{aligned}
\end{equation}

\end{appendices}

\end{document}